\documentclass[aps,prd,amssymb,nofootinbib,twocolumn,epsf]{revtex4}
\usepackage[usenames]{color} 
\usepackage{ulem} 
\usepackage{graphicx}
\usepackage{amssymb}
\usepackage{amsmath}
\usepackage{epstopdf}

\begin{document}


\title{Spectral Representations of Neutron-Star Equations of State}

\author{Lee Lindblom}

\affiliation{Theoretical Astrophysics 350-17, California Institute of
Technology, Pasadena, CA 91125}

\date{\today}
 
\begin{abstract}
Methods are developed for constructing spectral representations of
cold (barotropic) neutron-star equations of state.  These
representations are faithful in the sense that every physical equation
of state has a representation of this type, and conversely every such
representation satisfies the minimal thermodynamic stability criteria
required of any physical equation of state.  These spectral
representations are also efficient, in the sense that only a few
spectral coefficients are generally required to represent neutron-star
equations of state quiet accurately.  This accuracy and efficiency is
illustrated by constructing spectral fits to a large collection of
``realistic'' neutron-star equations of state.
\end{abstract}
 
\maketitle


\section{Introduction}
\label{s:introduction}

The gravitational field of a neutron star compresses the material in
its core to densities that exceed those inside normal atomic
nuclei.  The resulting matter is a mixture of free
baryons (neutrons and protons), leptons (electrons and muons), and
likely also smaller fractions of hyperons, mesons, or perhaps even
free quarks.  The basic thermodynamic relationship between the total
energy density $\epsilon$, and the pressure $p$ of this material is
called its equation of state, $\epsilon=\epsilon(p)$, and is
determined by the complicated micro-physical interactions between the
various particle species present in the mixture~\cite{shapiro83}.  

In addition to its dependence on the pressure, the energy density of a
mixture also depends typically on the temperature and other quantities
like the relative abundances of the different particle species.  Thus
there is no guarantee that a simple barotropic form,
$\epsilon=\epsilon(p)$, applies to any, let alone universally to all,
neutron-star matter.  Yet there is reason to expect that a universal
barotropic form might be an excellent approximation.  Neutron stars
are born when the cores of massive stars (and perhaps white dwarfs)
become unstable and undergo gravitational collapse.  Compression heats
the material as it collapses, to temperatures that exceed the binding
energies of all atomic nuclei.  Neutron-star matter always begins
therefore as a very hot plasma of free baryons and leptons, etc.  This
material is expected to evolve quickly to the lowest available energy
state as it cools by neutrino and photon emission, and this fixes the
relative abundances of the various particle species.  The thermal
energies of the particles fall rapidly below their Fermi levels, so
the thermal contribution to the energy rapidly becomes negligible.
The matter therefore is expected to evolve on a very short time scale
to a state that is well described by a temperature independent
barotropic equation of state.  This paper develops more efficient ways
to represent equations of state of this type.

The matter densities in the cores of neutron stars are well beyond the
reach of current laboratory experiments.  Heavy-ion scattering
(including most recently those conducted at RHIC and LHC) provides a
wealth of information about the interactions among the various
particles expected to make up neutron-star matter.  Unfortunately
those experiments bear only indirectly on the properties of the
equilibrium ground state, because the effective temperature of the
nuclear matter in the experiments is quite high.  Little insight is
provided therefore into effects, like complicated many body
interactions, that might play a role only in states with low
temperature and high density and pressure.

There has also been a significant effort over the past several decades
to understand this material from a theoretical perspective: hundreds
of papers devoted to modeling neutron-star matter have appeared in the
literature.  But the properties of this material are far outside the
realm where the usual arsenal of theoretical tools were designed to
work reliably, so it is not surprising that there is no consensus
among theoreticians yet on the neutron-star equation of state.  For
example, the current models' predictions of the pressure at a given
density still vary by about an order of magnitude~\cite{Read:2008iy}.
 
Direct observations of neutron stars may be the most promising
approach to understanding the properties of high-density nuclear
matter.  It is well known that the equation of state along with the
gravitational field equations determine the observable macroscopic
properties of neutron stars~\cite{Oppenheimer1939}, and conversely
that a complete knowledge of an appropriate set of macroscopic
properties (e.g. masses and radii) determines the equation of
state~\cite{Lindblom1992}.  Studies of neutron-star models show that
their macroscopic properties, like their masses and radii, vary widely
even within the current ``realistic'' class of equations of
state~\cite{ArnettBowers1977,Read:2008iy}.  So it has long been
recognized that accurate observations of the macroscopic properties of
neutron stars will provide significant constraints on the nuclear
equation of state.  Unfortunately the needed observations are quite
difficult to make.  Masses of several dozen neutron stars have now
been measured fairly accurately (see e.g. Ref.~\cite{Lattimer2007}),
and these observations have ruled out large classes of very ``soft''
equation of state models.  Only a few radius measurements have been
made however~\cite{Steiner2010}, and these are not reliable and
accurate enough yet to make solid quantitative measurements of the
equation of state itself possible.
   
There is reason to hope that more abundant and accurate measurements
of both neutron-star masses and radii will become available, however.
When the first accurate measurements are made, they are not likely to
be numerous and accurate enough to determine the entire high density
portion of the neutron-star equation of state.  Various attempts have
been made, therefore, to find representations of equations of state
that make their essential features depend on just a few parameters.
One approach is to use the parameters that characterize the nuclear
interaction models as a way to parametrize the equations of state
constructed from them.  These might include a number of micro-physical
parameters like the bulk nucleon incompressibility and symmetry energy
parameters in models of the nucleon interaction
potential~\cite{PAL1988}, or the coupling constants and mixing angles
in effective mean field theory descriptions~\cite{Glendenning1985}.
Comparing the masses and radii of neutron stars based
on these model equations of state should fix the values of the unknown
nuclear interaction model parameters.  This approach would clearly be
ideal if a reliable and accurate micro-physical model of neutron-star
matter were known.  Unfortunately there is no consensus that any of
the existing nuclear-matter models are good enough yet to describe
neutron-star matter accurately and reliably.
 
Another approach is to construct purely empirical fits rather than
micro-physics based models of the equation of state.  The first
attempts to do this~\cite{Lindblom1992, Vuille1995} approximated the
high density part of the equation of state as a simple polytrope,
i.e., an equation of state in which the pressure is proportional to a
power of the density.\footnote{The term relativistic polytrope is most
  commonly used for equations of state that satisfy
  $p\propto\rho^\Gamma$, where $\rho$ is the conserved rest-mass
  density and $\Gamma$ is the (constant) adiabatic index.  It is also
  used (less commonly) for those satisfying $p\propto \epsilon^\gamma$
  where $\epsilon$ is the total energy density and $\gamma$ is a
  constant.  The two definitions agree in the Newtonian limit.}  These
first simple fits were shown to reproduce the central pressures and
densities of neutron-star models based on ``realistic'' equations of
state with about 15\% accuracy~\cite{Lindblom1992}.  This type of
approximation can be improved by dividing the relevant range of
pressures into a number of intervals, $p_0< p_1< ... <
p_\mathrm{max}$, and fitting a different polytrope to the equation of
state in each interval.  Any level of accuracy can then be achieved by
using a sufficiently large number of intervals.  A number of authors
have proposed using piecewise polytropes to approximate the high
density part of the neutron-star equation of state~\cite{Vuille1995,
  VuilleIpser1999, Read:2008iy}, and these approximations turn out to
be quite efficient.  Fits of the high density parts of ``realistic''
neutron-star equations of state have been shown to achieve accuracies
of a few percent for piecewise polytropes with only a small number of
free parameters~\cite{Vuille1995, VuilleIpser1999, Read:2008iy}.  The
most extensive study to date uses fits with four free parameters that
give average errors of only a few percent for 34 realistic equations
of state~\cite{Read:2008iy}.  Other types of empirical fits to
neutron-star equations of state have also been reported in the
literature~\cite{Haensel2004,Shibata2005, Shibata2006,Zdunik2006}.
These provide high accuracy approximations of particular realistic
equations of state (generally using fifteen to twenty parameters to do
this), and do not appear to have been intended as efficient ways to
model large classes of equations of state.

This paper continues the effort to construct efficient empirical
representations of ``realistic'' neutron-star equations of state.  New
methods are described here for constructing parametric representations
based on spectral fits.  Spectral representations are generalizations
of the Fourier series used to represent periodic functions.  It is
shown in Sec.~\ref{s:SpectralEOS} that spectral representations can be
constructed that are faithful, in the sense that every physical
equation of state has such a representation and conversely that every
such representation satisfies the basic thermodynamic stability
conditions required of any equation of state.  It is also shown in
Sec.~\ref{s:RealisticEOS} that these spectral representations do a
good job of representing the currently available ``realistic''
neutron-star equations of state.  A suitably constructed two-parameter
spectral representation is shown to be about as accurate as the most
carefully studied four-parameter polytrope fits~\cite{Read:2008iy}.
For smooth equations of state, the errors in the spectral fits
decrease exponentially as the number of parameters is increased, while
piecewise polytrope fits generally decrease only quadratically.  These
spectral fits make it possible therefore to provide very accurate
representations of the neutron-star equation of state using only a
small number parameters.  They should provide an important new tool
for extracting the high density equation of state from neutron-star
observations.


\section{Spectral Representations of the
Equation of State}
\label{s:SpectralEOS}

Any equation of state, $\epsilon=\epsilon(p)$, can be represented in a
``spectral'' expansion, as linear combinations of basis functions
$\Phi_k(p)$:
\begin{eqnarray}
\epsilon(p)=\sum_k \epsilon_k \Phi_k(p).
\label{e:EpsilonSpectralExpansion}
\end{eqnarray}
Any complete set of functions, such as the Fourier basis functions or
the Chebyshev polynomials, could be used as the $\Phi_k$ in these
expansions.  Equations of state are determined in such representations
by their spectral coefficients, $\epsilon_k$.  Truncated versions of
these expansions, in which only a finite number of terms are kept,
provide approximate parametric representations of arbitrary equations
of state: $\epsilon =\epsilon(p,\epsilon_k)$.

Physical equations of state must be non-negative, $\epsilon(p)\geq 0$,
and monotonically increasing functions, $d\epsilon(p)/dp\geq 0$, to
insure thermodynamic stability.  Since almost all functions fail to
satisfy these conditions, it follows that almost all choices of
spectral coefficients, $\epsilon_k$, in an expansion such as
Eq.~(\ref{e:EpsilonSpectralExpansion}) represent functions that can
not be equations of state.  Thus the spectral coefficients
$\epsilon_k$ obtained by fitting to a physical equations of state are
likely to produce a representation that violates basic thermodynamic
stability.  So unfortunately, representing an equation of state with a
straightforward spectral expansion is not particularly useful.

Instead, faithful representations are needed: ones that ensure the
positivity and monotonicity conditions for every choice of spectral
coefficients.  Methods of constructing faithful representations of the
equation of state are presented in the following sections.  Spectral
representations of the standard form of the equation of state, in
which the energy density is expressed as a function of the pressure
$\epsilon=\epsilon(p)$, are presented in
Sec.~\ref{s:PressureBasedForms}.  For some applications it
is more convenient to express the equation of state in terms of the
relativistic enthalpy $\epsilon=\epsilon(h)$.  Spectral
representations of these enthalpy based forms are given in
Sec.~\ref{s:EnthalpyBasedForms}.


\subsection{Pressure Based Forms}
\label{s:PressureBasedForms}

An equation of state $\epsilon(p)$ determines, and is determined by
(up to an integration constant), the adiabatic index $\Gamma(p)$,
defined by
\begin{eqnarray}
\Gamma(p) = \frac{\epsilon+p}{p}\frac{dp}{d\epsilon}.
\label{e:GammaDef}
\end{eqnarray}
Given $\Gamma(p)$, the equation of state $\epsilon(p)$ is determined
simply by integrating the first-order ordinary differential equation,
\begin{eqnarray}
 \frac{d\epsilon(p)}{dp}= \frac{\epsilon(p)+p}{p\,\Gamma(p)}.
\label{e:EOSode}
\end{eqnarray}
The adiabatic index must be positive $\Gamma(p)>0$ to ensure
thermodynamic stability, but it need not be monotonic.  Thus a
larger class of functions represent possible physical adiabatic
indices, and this makes it easier to represent equations of state
through spectral expansions of $\Gamma(p)$.  In particular, every
physical equation of state can be represented by the following
spectral expansion of the adiabatic index $\Gamma(p)$:
\begin{eqnarray}
\Gamma(p) = \exp \left[\sum_k \gamma_k \Phi_k(p)\right].
\end{eqnarray}
Conversely, every choice of $\gamma_k$ (for which the series in this
expansion converges) results in a positive adiabatic index, and thus
an equation of state from Eq.~(\ref{e:EOSode}), that satisfies the
positivity and thermodynamic stability conditions.  So this representation
is faithful in the sense defined in Sec.~\ref{s:SpectralEOS}.

The construction of an explicit spectral representation of the
equation of state requires a choice for the basis functions
$\Phi_k(p)$.  To that end, it is useful to define a dimensionless
logarithmic pressure variable:
\begin{eqnarray}
x = \log(p/p_0).
\end{eqnarray}
The constant $p_0$ is a scale factor, chosen here to be the minimum
value of the pressure, $p_0\leq p$, in the domain where the spectral
expansions are to be used.  The following expansion of the adiabatic
index $\Gamma(x)$ is found to be useful and effective,
\begin{eqnarray}
\Gamma(x) = \exp\left(\sum_k \gamma_k \,x^k\right).
\label{e:GammaX}
\end{eqnarray}
One advantage of this simple power-law basis is that the lowest order
spectral coefficients $\gamma_k$ have fairly simple physical
interpretations.  For example the lowest order coefficient,
$\gamma_0$, is determined by the adiabatic index evaluated at the
reference pressure: $\gamma_0 = \log \Gamma(p_0)$.  Similarly, the
next coefficient, $\gamma_1$, determines the behavior of the adiabatic
index, $\Gamma(p)\approx \Gamma(p_0)(p/p_0)^{\gamma_1}$ for pressures
near $p_0$, i.e. for $x=\log(p/p_0)\ll 1$.  

The power law basis, $\Phi_k(x)=x^k$, has the advantage of simplicity.
While it might be advantageous to choose another basis like the
Chebyshev polynomials for some purposes, these advantages can only be
fully exploited by using a knowledge of the exact range, $0\leq x\leq
x_{\mathrm{max}}$, and re-scaling $x$ in the optimal way.  The
additional information, like $x_{\mathrm{max}}$ for example, needed to
do that will not be available {\it a priori} for the real neutron-star
equation of state.  So here the simple power law basis is used, and
fortunately this choice seems to work quite well.

Given an adiabatic index, $\Gamma(p)$, it is straightforward to
determine the equation of state, $\epsilon(p)$, by integrating the
ordinary differential equation, Eq.~(\ref{e:EOSode}).  The solutions
and hence the equation of state can be reduced to quadratures:
\begin{eqnarray}
\epsilon(p)= \frac{\epsilon_0}{\mu(p)}+
\frac{1}{\mu(p)}\int_{p_0}^p\frac{\mu(p')}{\Gamma(p')}dp',
\label{e:EpsilonX}
\end{eqnarray}
where $\mu(p)$ is defined as
\begin{eqnarray}
\mu(p)=\exp\left[-\int_{p_0}^p \frac{dp'}{p'\,\Gamma(p')}\right],
\end{eqnarray}
and where $\epsilon_0=\epsilon(p_0)$ is the constant of integration
needed to fix the solution.  This $\epsilon_0$ is fixed in the fits
performed in Sec.~\ref{s:RealisticEOS} by matching to a low density
equation of state at the pressure, $p_0$ ({\it i.e.} at $x_0=0$),
chosen to be a point somewhat below nuclear density.

The quadratures indicated in Eq.~(\ref{e:EpsilonX}) can not be done
analytically for the expansion given in Eq.~(\ref{e:GammaX}), so an
explicit analytic expression for the equation of state is not
available in this case.  However, the integrands in these quadratures
are analytic functions that can be integrated numerically very
accurately and efficiently.  Using Gaussian quadrature for example,
double precision accuracy can be achieved using about 10 points for
each integral.  So there is very little practical difference between
having an explicit analytic expression for the equation of state, and
the expression in Eq.~(\ref{e:EpsilonX}) in terms of quadratures of
explicit analytic functions.

It might be advantageous in some situations to construct
spectral expansions for the equation of state using thermodynamic
quantities other than $\Gamma(p)$.  For example,
the adiabatic sound speed, $v(p)$, defined by
\begin{eqnarray}
v^2(p)=c^2\left[\frac{d\epsilon(p)}{dp}\right]^{-1},
\end{eqnarray}
(where $c$ is the speed of light) could be used to obtain the equation
of state by integrating the simple ordinary differential equation,
\begin{eqnarray}
\frac{d\epsilon(p)}{dp}=\frac{c^2}{v^2(p)}.
\end{eqnarray}
The thermodynamic stability condition, $0\leq v^2$, could be enforced
in this case by constructing the following spectral expansion,
\begin{eqnarray}
v^2(p)=c^2 \exp\left[\sum_k v_k \Phi_k(p)\right].
\end{eqnarray}
Alternatively, it might be desirable to enforce both the thermodynamic
stability condition and the ``causality'' conditions,\footnote{The
  condition $v^2\leq c^2$ only represents a true causality condition
  if the equation of state $\epsilon(p)$ describes both the time
  dependent and the equilibrium properties of the material.  In
  neutron-star matter, various strong and weak nuclear interactions
  determine the relative abundances of the various particle species in
  the equilibrium state.  Thus $v^2$ evaluated for the equilibrium
  equation of state only describes the sound propagation speed for low
  enough frequency waves that the material remains continuously in
  equilibrium.  The condition $v^2\leq c^2$ may or may not represent a
  causality constraint therefore on sound waves with short enough
  wavelengths to be physically relevant in neutron stars.} $0\leq v^2
\leq c^2$, by constructing the spectral expansion in the following
way,
\begin{eqnarray}
v^2(p)=c^2\left\{1+\exp\left[-\sum_k v_k \Phi_k(p)\right]\right\}^{-1}.
\end{eqnarray}

For the remainder of this paper, spectral representations of the
equation of state will be based on the familiar adiabatic index
$\Gamma(p)$.  The discussion in Sec.~\ref{s:RealisticEOS} shows that
$\Gamma(p)$ is a reasonably slowly varying function for ``realistic''
neutron-star equations of state, which can be represented fairly
accurately using expansions having only a few terms.  Using
$\Gamma(p)$ for these expansions also allows us to make
straightforward comparisons with published piecewise-polytrope
approximations to the equation of state~\cite{Read:2008iy}.  The
accuracy and efficiency of these expansions in representing
``realistic'' neutron-star equations of state is explored in
Sec.~\ref{s:RealisticEOS}.


\subsection{Enthalpy Based Forms}
\label{s:EnthalpyBasedForms}

The spectral expansions of the standard representation of equation of
state, $\epsilon=\epsilon(p)$, should be quite useful for many
applications.  For some applications, however, the standard
representation, $\epsilon=\epsilon(p)$, is not ideal.  For example, a
useful form of the relativistic stellar structure
equations~\cite{Lindblom1992} requires the equation of state to be
expressed in terms of the relativistic enthalpy, $h$.  For
applications such as this, $\epsilon=\epsilon(p)$ must be re-written
as a pair of equations $\epsilon=\epsilon(h)$ and $p=p(h)$, where $h$
is defined as
\begin{eqnarray}
h(p) = \int_0^p \frac{dp'}{\epsilon(p')+ p'}.
\label{e:EnthalpyDef}
\end{eqnarray} 
The needed expressions, $\epsilon=\epsilon(h)$ and $p=p(h)$, are
constructed by inverting $h=h(p)$ from Eq.~(\ref{e:EnthalpyDef}) to
obtain $p=p(h)$, and composing the result with the standard equation of
state, $\epsilon=\epsilon(p)$, to obtain $\epsilon(h)=\epsilon[p(h)]$.

The transformations needed to construct $\epsilon=\epsilon(h)$ and
$p=p(h)$ are difficult to perform numerically in an efficient and
accurate way.  Therefore it may be preferable to construct a spectral
expansion of the equation of state based directly on $h$.  This can be
done using the methods described above for the standard
$\epsilon=\epsilon(p)$ representation.  To do this a spectral
expansion of the adiabatic index, considered now as a function of the
enthalpy $\Gamma(h)$, must be defined.  The scaled enthalpy variable
$x = \log(h/h_0)$, is found to be useful, where $h_0$ is the lower
bound on the enthalpy, $h_0\leq h$, in the domain where the spectral
expansions are constructed.  The expression for $\Gamma(x)$ given in
Eq.~(\ref{e:GammaX}) then provides a useful expansion for $\Gamma(h)$:
\begin{eqnarray}
\Gamma(h) = \exp\left\{\sum_k \gamma_k \,\left[\log\left(\frac{h}{h_0}
\right)\right]^k\right\}.
\label{e:GammaH}
\end{eqnarray}
Next, the functions $p(h)$ and $\epsilon(h)$ are defined by the system
of ordinary differential equations,
\begin{eqnarray}
\frac{dp}{dh} &=& \epsilon + p,\\
\frac{d\epsilon}{dh} &=& \frac{(\epsilon + p)^2}{p\,  \Gamma(h)},
\end{eqnarray}
that follow from the definitions of $h$, Eq.~(\ref{e:EnthalpyDef}),
and $\Gamma$, Eq.~(\ref{e:GammaDef}).  The general solution to these
equations can be reduced to quadrature:
\begin{eqnarray}
p(h)&=&p_0 \exp\left[\int_{h_0}^h \frac{e^{h'}dh'}{\tilde \mu(h')}
\right],\label{e:PressueH}\\
\epsilon(h)&=& p(h)  \frac{e^h -\tilde \mu(h)}{\tilde \mu(h)},
\label{e:EnthalpyH}
\end{eqnarray}
where $\tilde \mu(h)$ is defined as,
\begin{eqnarray}
\tilde \mu(h) = \frac{p_0\, e^{h_0}}{\epsilon_0 + p_0} 
+ \int_{h_0}^h \frac{\Gamma(h')-1}{\Gamma(h')} e^{h'}dh'.
\label{e:TildeMuDef}
\end{eqnarray}
The constants $p_0$ and $\epsilon_0$ are defined by $p_0=p(h_0)$ and
$\epsilon_0=\epsilon(h_0)$ respectively.  While these quadratures can
not be done analytically for the spectral expansion defined in
Eqs.~(\ref{e:GammaH}), they can be done numerically very efficiently
and accurately using Gaussian quadrature, as in the standard equation
of state case.

 
\section{Spectral Fits to Realistic Equations of State}
\label{s:RealisticEOS}

The discussion in Sec.~\ref{s:SpectralEOS} shows how any equation of
state can be represented by spectral expansions of the adiabatic
index, like the one given in Eq.~(\ref{e:GammaX}).  When these
expansions are truncated, keeping only a finite number of terms, they
produce fits to the equation of state, $\epsilon_{\mathrm{fit}}(p)$
that are expected to converge to the exact $\epsilon(p)$ as the number
of terms in the expansion increases.  In analogy with Fourier series,
the rate of convergence for these fits should be exponential for
smooth equations of state, and power law for less than smooth cases.
The smoothness of an equation of state is determined by the details of
the micro-physics that controls the properties of the material.  The
convergence rate of an expansion will be reduced therefore, from
exponential to power law, when phase transitions or other non-smooth
transitions are present.  The number of terms required to achieve a
certain level of accuracy in $\epsilon_{\mathrm{fit}}(p)$, therefore,
will depend on the smoothness and variability of the adiabatic index
$\Gamma(p)$, and the suitability of the chosen spectral basis
functions $\Phi_k(p)$.

The accuracy and practicality of spectral expansions for two forms of
the equation of state, $\epsilon=\epsilon(p)$ and
$\epsilon=\epsilon(h)$, are evaluated in this section by constructing
fits to 34 ``realistic'' neutron-star equations of state.  These
spectral fits are based on finite spectral expansions of $\Gamma(p)$
and $\Gamma(h)$ respectively.  The equations of state used for these
fits are the same as those used by Read, Lackey, Owen and
Friedman~\cite{Read:2008iy} in their study of piecewise-polytrope
approximations.  These realistic equations of state are based on a
variety of different models for the composition of neutron-star
matter, and a variety of different models for the interactions between
the particle species present in the model material.  Descriptions of
these realistic equation of state models, and references to the
original publications on each of these equations of state are given in
Ref.~\cite{Read:2008iy}, and are not repeated here.  The individual
equations of state are referred to here using the abbreviations used
in Ref.~\cite{Read:2008iy}, {\it e.g.}  PAL6, APR1, BGN1H1, etc.  The
list of these equations of state are given in the first column of
Table III of Ref.~\cite{Read:2008iy}, and the first columns of
Tables~\ref{t:TableI} and \ref{t:TableII} in this paper.
\begin{table*}[!htb]
\begin{center}
\caption{Spectral Expansions of the Standard $\epsilon=\epsilon(p)$ Form of
  Realistic Neutron-Star Equations of State
\label{t:TableI}}
\begin{tabular}{|l|c|cccc|cccc|ccc|}
\hline\hline
    EOS   &$\Delta_{P4}$ &$\Delta_{S2}$ &$\Delta_{S3}$ &$\Delta_{S4}$ 
     &$\Delta_{S5}$ &$\gamma_0$ &$\gamma_1$
    &$\gamma_2$ &$\gamma_3$ &$p_0$ &$\epsilon_0/c^2$ &$x_{\mathrm{max}}$\\
\hline
    PAL6 &  0.0076 &  0.0025 &  0.0014 &  0.0005 &  0.0001 &  0.8622 & -0.0677 &  0.0181 & -0.0017 & $    3.01\times 10^{33}$ & $    2.03\times 10^{14}$  &    5.79\\
     SLy &  0.0208 &  0.0076 &  0.0028 &  0.0010 &  0.0002 &  0.9865 &  0.1110 & -0.0301 &  0.0022 & $    1.64\times 10^{33}$ & $    2.05\times 10^{14}$  &    6.73\\
     AP1 &  0.0831 &  0.0552 &  0.0199 &  0.0098 &  0.0029 &  0.4132 &  0.5594 & -0.1270 &  0.0085 & $    1.16\times 10^{33}$ & $    2.02\times 10^{14}$  &    7.38\\
     AP2 &  0.0411 &  0.0236 &  0.0091 &  0.0033 &  0.0017 &  0.7065 &  0.2866 & -0.0659 &  0.0047 & $    1.34\times 10^{33}$ & $    2.02\times 10^{14}$  &    7.27\\
     AP3 &  0.0352 &  0.0204 &  0.0026 &  0.0011 &  0.0009 &  0.9214 &  0.2097 & -0.0383 &  0.0019 & $    1.51\times 10^{33}$ & $    2.02\times 10^{14}$  &    6.87\\
     AP4 &  0.0273 &  0.0194 &  0.0016 &  0.0015 &  0.0013 &  0.8651 &  0.1548 & -0.0151 & -0.0002 & $    1.50\times 10^{33}$ & $    2.02\times 10^{14}$  &    7.04\\
     FPS &  0.0120 &  0.0045 &  0.0037 &  0.0034 &  0.0021 &  1.1561 & -0.0468 &  0.0081 & -0.0010 & $    1.19\times 10^{33}$ & $    2.04\times 10^{14}$  &    7.04\\
    WFF1 &  0.0320 &  0.0387 &  0.0067 &  0.0066 &  0.0062 &  0.6785 &  0.2626 & -0.0215 & -0.0008 & $    1.14\times 10^{33}$ & $    2.04\times 10^{14}$  &    7.48\\
    WFF2 &  0.0342 &  0.0200 &  0.0077 &  0.0051 &  0.0041 &  0.8079 &  0.2680 & -0.0558 &  0.0039 & $    1.32\times 10^{33}$ & $    2.04\times 10^{14}$  &    7.25\\
    WFF3 &  0.0314 &  0.0081 &  0.0081 &  0.0060 &  0.0059 &  1.4126 & -0.1797 &  0.0389 & -0.0035 & $    0.81\times 10^{33}$ & $    2.03\times 10^{14}$  &    7.32\\
    BBB2 &  0.1016 &  0.0283 &  0.0238 &  0.0167 &  0.0042 &  0.7390 &  0.4555 & -0.1406 &  0.0121 & $    1.37\times 10^{33}$ & $    2.05\times 10^{14}$  &    6.97\\
  BPAL12 &  0.0739 &  0.0138 &  0.0086 &  0.0043 &  0.0018 &  1.1081 & -0.3078 &  0.0891 & -0.0081 & $    2.51\times 10^{33}$ & $    2.05\times 10^{14}$  &    6.08\\
     ENG &  0.0527 &  0.0181 &  0.0168 &  0.0138 &  0.0118 &  0.9820 &  0.2716 & -0.0862 &  0.0075 & $    1.33\times 10^{33}$ & $    2.04\times 10^{14}$  &    6.98\\
    MPA1 &  0.0365 &  0.0223 &  0.0022 &  0.0022 &  0.0019 &  1.0215 &  0.1653 & -0.0235 & -0.0004 & $    1.51\times 10^{33}$ & $    2.04\times 10^{14}$  &    6.63\\
     MS1 &  0.0581 &  0.0256 &  0.0052 &  0.0039 &  0.0004 &  0.9189 &  0.1432 &  0.0122 & -0.0094 & $    4.54\times 10^{33}$ & $    2.04\times 10^{14}$  &    5.05\\
     MS2 &  0.0155 &  0.0074 &  0.0013 &  0.0005 &  0.0001 &  0.9598 & -0.0527 &  0.0091 & -0.0035 & $    4.10\times 10^{33}$ & $    2.04\times 10^{14}$  &    4.77\\
    MS1b &  0.0206 &  0.0179 &  0.0058 &  0.0032 &  0.0004 &  1.2132 & -0.0648 &  0.0561 & -0.0111 & $    3.18\times 10^{33}$ & $    2.03\times 10^{14}$  &    5.40\\
      PS &  0.0568 &  0.0566 &  0.0294 &  0.0284 &  0.0182 &  1.3896 & -0.8472 &  0.2636 & -0.0218 & $    5.49\times 10^{33}$ & $    2.05\times 10^{14}$  &    4.77\\
     GS1 &  0.0536 &  0.0762 &  0.0385 &  0.0333 &  0.0265 &  1.8662 & -1.4266 &  0.4450 & -0.0389 & $    3.22\times 10^{33}$ & $    2.04\times 10^{14}$  &    6.28\\
     GS2 &  0.0416 &  0.0582 &  0.0427 &  0.0425 &  0.0294 &  1.4580 & -0.7219 &  0.1828 & -0.0117 & $    4.06\times 10^{33}$ & $    2.04\times 10^{14}$  &    4.99\\
  BGN1H1 &  0.0435 &  0.0792 &  0.0460 &  0.0439 &  0.0328 &  1.3450 & -0.0996 & -0.0833 &  0.0161 & $    2.30\times 10^{33}$ & $    2.05\times 10^{14}$  &    6.40\\
    GNH3 &  0.0092 &  0.0130 &  0.0090 &  0.0081 &  0.0057 &  1.0366 & -0.0044 & -0.0440 &  0.0075 & $    3.77\times 10^{33}$ & $    2.06\times 10^{14}$  &    5.26\\
      H1 &  0.0226 &  0.0200 &  0.0117 &  0.0089 &  0.0069 &  1.0653 &  0.0362 & -0.1098 &  0.0179 & $    3.17\times 10^{33}$ & $    2.04\times 10^{14}$  &    5.04\\
      H2 &  0.0300 &  0.0181 &  0.0133 &  0.0072 &  0.0069 &  1.0743 &  0.2250 & -0.2029 &  0.0290 & $    2.96\times 10^{33}$ & $    2.04\times 10^{14}$  &    4.98\\
      H3 &  0.0308 &  0.0130 &  0.0109 &  0.0086 &  0.0066 &  1.1340 &  0.0925 & -0.1303 &  0.0190 & $    3.12\times 10^{33}$ & $    2.04\times 10^{14}$  &    4.91\\
      H4 &  0.0098 &  0.0098 &  0.0097 &  0.0069 &  0.0063 &  1.0526 &  0.1695 & -0.1200 &  0.0150 & $    3.16\times 10^{33}$ & $    2.04\times 10^{14}$  &    5.21\\
      H5 &  0.0214 &  0.0150 &  0.0126 &  0.0054 &  0.0054 &  1.0106 &  0.2765 & -0.2011 &  0.0270 & $    2.97\times 10^{33}$ & $    2.04\times 10^{14}$  &    5.09\\
      H6 &  0.0185 &  0.0137 &  0.0133 &  0.0130 &  0.0100 &  1.0650 & -0.0196 & -0.0474 &  0.0077 & $    3.35\times 10^{33}$ & $    2.04\times 10^{14}$  &    4.80\\
      H7 &  0.0139 &  0.0132 &  0.0103 &  0.0059 &  0.0056 &  0.9582 &  0.1619 & -0.1294 &  0.0177 & $    3.19\times 10^{33}$ & $    2.04\times 10^{14}$  &    5.22\\
    PCL2 &  0.0227 &  0.0252 &  0.0121 &  0.0090 &  0.0075 &  1.0410 &  0.0173 & -0.0904 &  0.0150 & $    2.90\times 10^{33}$ & $    2.04\times 10^{14}$  &    5.44\\
    ALF1 &  0.0947 &  0.0669 &  0.0453 &  0.0369 &  0.0305 &  1.0143 & -0.3102 &  0.1809 & -0.0248 & $    1.50\times 10^{33}$ & $    2.05\times 10^{14}$  &    6.21\\
    ALF2 &  0.0655 &  0.0629 &  0.0450 &  0.0256 &  0.0230 &  0.4613 &  1.5237 & -0.5817 &  0.0571 & $    1.50\times 10^{33}$ & $    2.05\times 10^{14}$  &    5.79\\
    ALF3 &  0.0371 &  0.0355 &  0.0139 &  0.0132 &  0.0131 &  0.8536 &  0.2405 & -0.0743 &  0.0041 & $    1.50\times 10^{33}$ & $    2.05\times 10^{14}$  &    6.14\\
    ALF4 &  0.0453 &  0.0652 &  0.0166 &  0.0089 &  0.0088 &  0.8806 &  0.0656 &  0.0765 & -0.0177 & $    1.50\times 10^{33}$ & $    2.05\times 10^{14}$  &    5.97\\
\hline
 Average &  0.0383 &  0.0287 &  0.0149 &  0.0114 &  0.0085 &&&&&&&\\
\hline\hline
 \end{tabular}
\end{center}
\end{table*}

Approximate equations of state, $\epsilon_{\mathrm{fit}}(p)$, have
been constructed for each of the realistic neutron-star equations of
state listed in Table~\ref{t:TableI}.  These approximations are based
on Eq.~(\ref{e:EpsilonX}) with $\Gamma(x)$ determined by the spectral
expansion in Eq.~(\ref{e:GammaX}).  The $\epsilon_{\mathrm{fit}}(p)$
constructed in this way depend on the pressure through the variable
$x=\log(p/p_0)$, as well as the spectral coefficients $\gamma_k$:
$\epsilon_{\mathrm{fit}}=\epsilon_{\mathrm{fit}}(x,\gamma_k)$.  The
optimal choice of spectral coefficients, $\gamma_k$, is made by
minimizing the differences between
$\epsilon_{\mathrm{fit}}(x_i,\gamma_k)$ and the exact
$\epsilon_i=\epsilon(x_i)$ for a set of pressures, $x_i$, from the
realistic neutron-star equation of state tables.  These differences
are measured by constructing the residual:
\begin{eqnarray}
\Delta^2(\gamma_k)=\sum_{i=1}^N\frac{1}{N}\left\{
\log\left[\frac{\epsilon_{\mathrm{fit}}(x_i,\gamma_k)}
{\epsilon_i}\right]\right\}^2,
\label{e:ResidualDef}
\end{eqnarray}
where the sum is over all the pressures in the tabulated realistic
equation of state in the range $p_0\leq p_i \leq p_{\mathrm{max}}$.
These are the pressures that may be present in the cores of neutron
stars where the equation of state is not well known: $p_0$ is the
pressure where the baryon density is $\rho_0=2\times 10^{14}\,g/cm^3$,
and $p_{\mathrm{max}}$ is the central pressure of the maximum-mass
non-rotating neutron-star model for the particular equation of state.
The constants $p_0$, $x_{\mathrm{max}}=\log(p_{\mathrm{max}}/p_0)$,
and the total energy density, $\epsilon_0=\epsilon(p_0)$, are given
(in cgs units) in the last three columns of Table~\ref{t:TableI} for
each of the realistic equations of state.  This range of pressures
coincides with the range used by Read, Lackey, Owen and
Friedman~\cite{Read:2008iy} to construct their piecewise-polytrope
fits.  Using the same range of pressures here makes comparison with
their work more straightforward.
\begin{figure}[t]
\centerline{\includegraphics[width=3in]{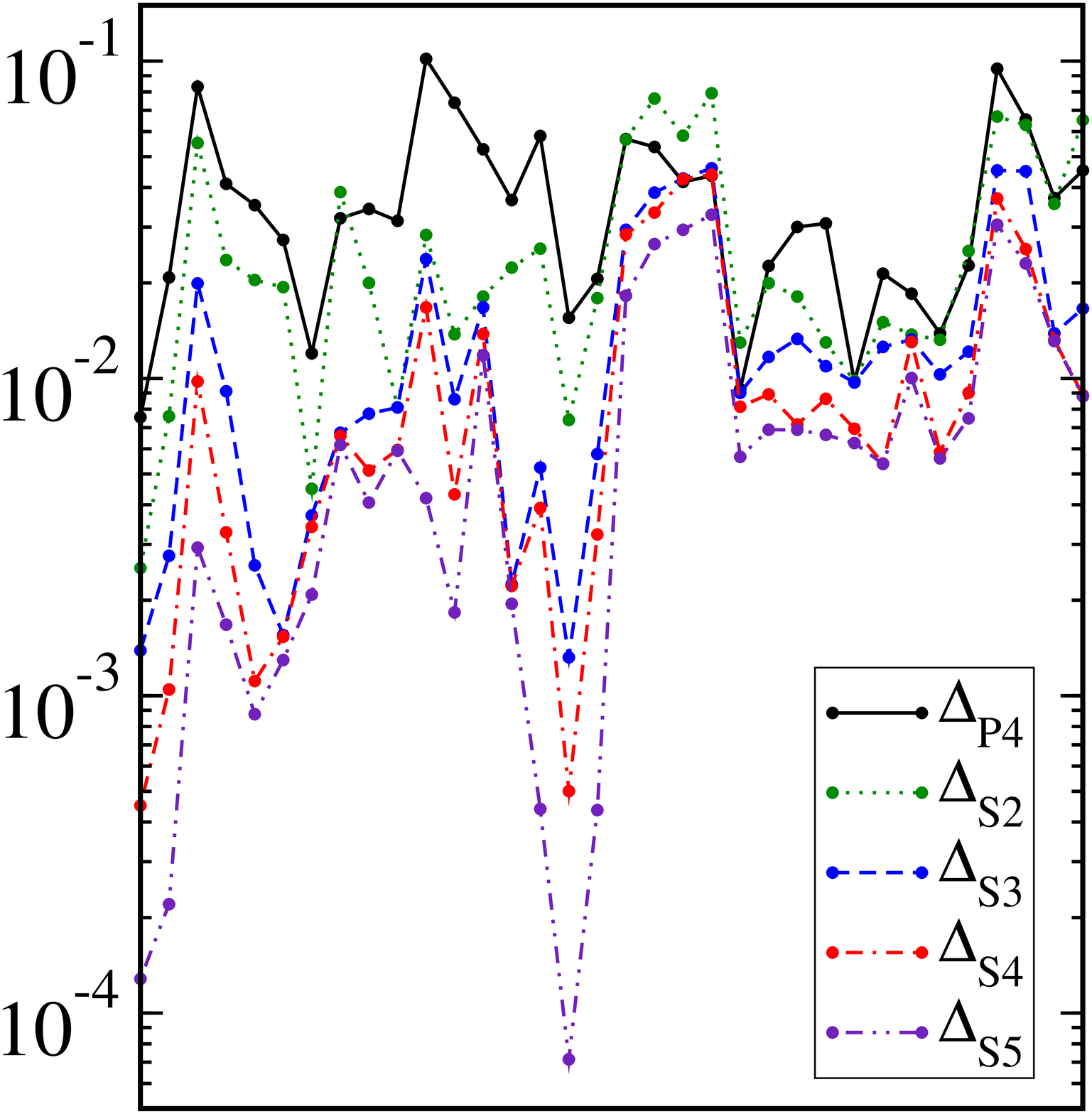}}
\caption{\label{f:chisqr} Residuals $\Delta_{Sk}$ are illustrated for
  several spectral fits and for the polynomial fit $\Delta_{P4}$ for
  each of the realistic equation of state models, which are
  represented as points along the horizontal axis in this figure.
  These residuals are for standard pressure-based representations of
  the equation of state, $\epsilon=\epsilon(p)$, which are also given
  in Table~\ref{t:TableI}. }
\end{figure}

The spectral coefficients, $\gamma_k$, that determine the particular
$\epsilon_{\mathrm{fit}}(x,\gamma_k)$ are chosen to minimize the
residual $\Delta(\gamma_k)$ for each realistic equation of state.  The
minimization process was carried out with an algorithm based on the
Levenberg-Marquardt method~\cite{numrec_f}, starting with initial
estimates, $\gamma_0=1$ and $\gamma_k=0$ for $k\geq 1$.  Approximate
$\epsilon_{\mathrm{fit}}$ were constructed in this way for expansions
containing 2, 3, 4, and 5 spectral basis functions.  The minimum
values of the residuals, $\Delta_{S2}$, $\Delta_{S3}$, $\Delta_{S4}$,
and $\Delta_{S5}$, for these cases are listed for each equation of
state in Table~\ref{t:TableI}.  The average values of these minimum
residuals decrease from about $2.9\%$ for the 2-parameter fits, to
about $0.9\%$ for the 5-parameter fits.  For comparison the residuals
$\Delta_{P4}$ for the 4-parameter piecewise-polytrope fits of Reid,
Lackey, Owen, and Friedman~\cite{Read:2008iy} are also given in
Table~\ref{t:TableI} for each equation of state.\footnote{The
  residuals reported in Table III of Ref.~\cite{Read:2008iy} differ
  from the $\Delta_{P4}$ residual listed in Table~\ref{t:TableI} in
  two ways.  The first difference is the residuals reported in
  Ref.~\cite{Read:2008iy} are evaluated using base-10 logarithms,
  rather than the natural logarithms used here.  This difference makes
  the Ref.~\cite{Read:2008iy} residuals smaller by the factor
  $\log_{10}e\approx 2.3$.  The second difference is that
  $\Delta_{P4}$ reported here measures the accuracy of the
  piecewise-polytrope fits for $\epsilon(p)$, while the residuals
  reported in Ref.~\cite{Read:2008iy} measure the accuracy of those
  fits for $p(\rho)$ where $\rho$ is the baryon density of the
  material. }  The values of the residuals, $\Delta_{S2}$,
$\Delta_{S3}$, $\Delta_{S4}$, $\Delta_{S5}$, and $\Delta_{P4}$, are
also shown graphically in Fig.~\ref{f:chisqr}.  Points along the
horizontal axis in Fig.~\ref{f:chisqr} represent the different
realistic equations of state in the order listed in
Table~\ref{t:TableI}.  These results show that the spectral fits are
convergent, and do a fairly good job of approximating this collection
of realistic equations of state.  The 2-parameter spectral fits have
smaller residuals than the 4-parameter piecewise-polytrope fits for
most of these equations of state.  Also listed in Table~\ref{t:TableI}
are the optimal values of the spectral coefficients, $\gamma_0$,
$\gamma_1$, $\gamma_2$, and $\gamma_3$, for the 4-parameter spectral
approximation to each equation of state.  These values, together with
the tabulated constants $\epsilon_0$ and $p_0$ can be used to
re-construct the complete 4-parameter spectral fits for $\epsilon(p)$,
using Eqs.~(\ref{e:GammaX}) and (\ref{e:EpsilonX}).
\begin{figure}[t]
\centerline{\includegraphics[width=3in]{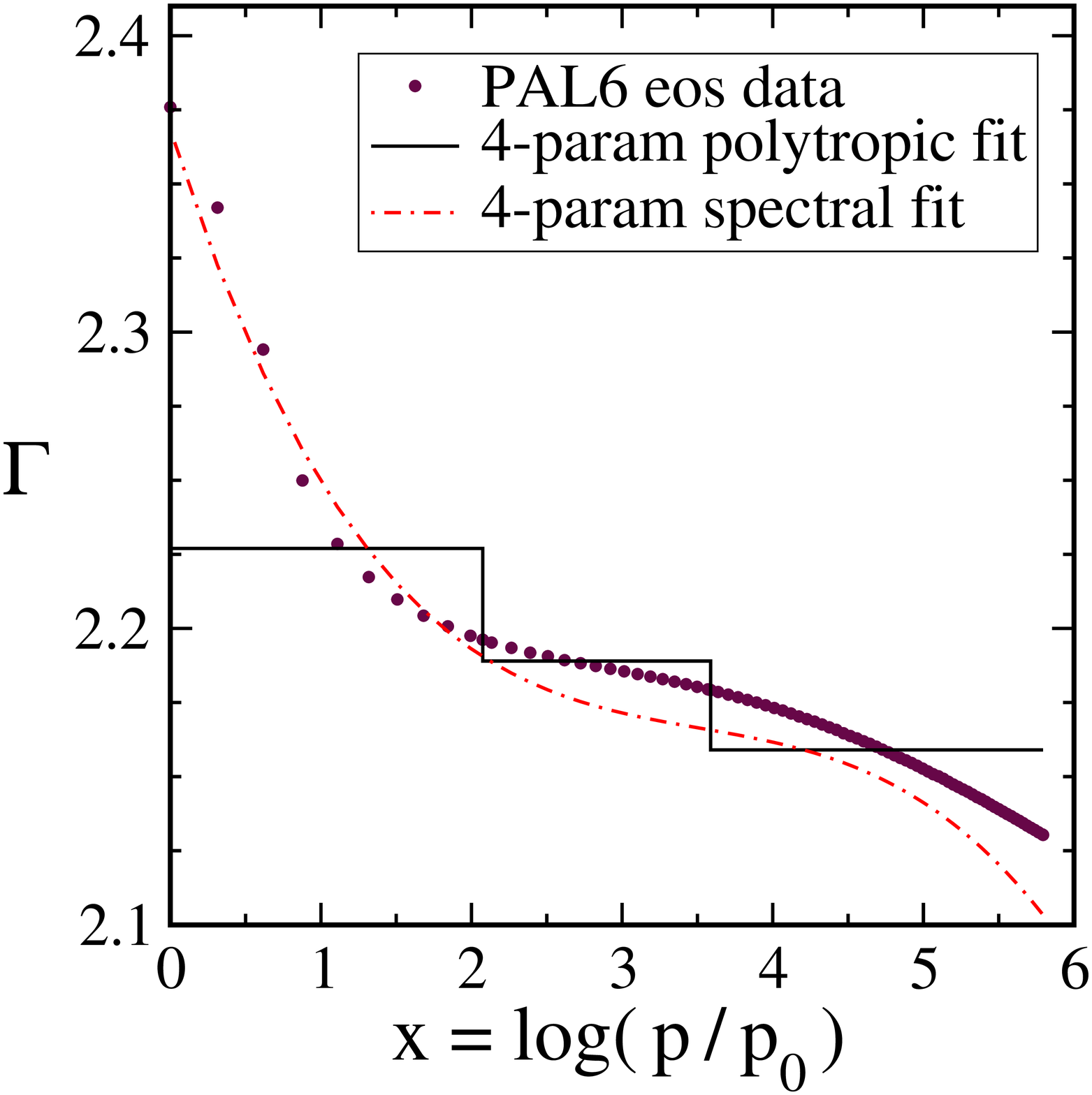}}
\caption{\label{f:pal6_gamma} Adiabatic index as a
function of pressure, $\Gamma(x)$, for various fits to the PAL6
equation of state.}
\end{figure}

\begin{figure}[t]
\centerline{\includegraphics[width=3in]{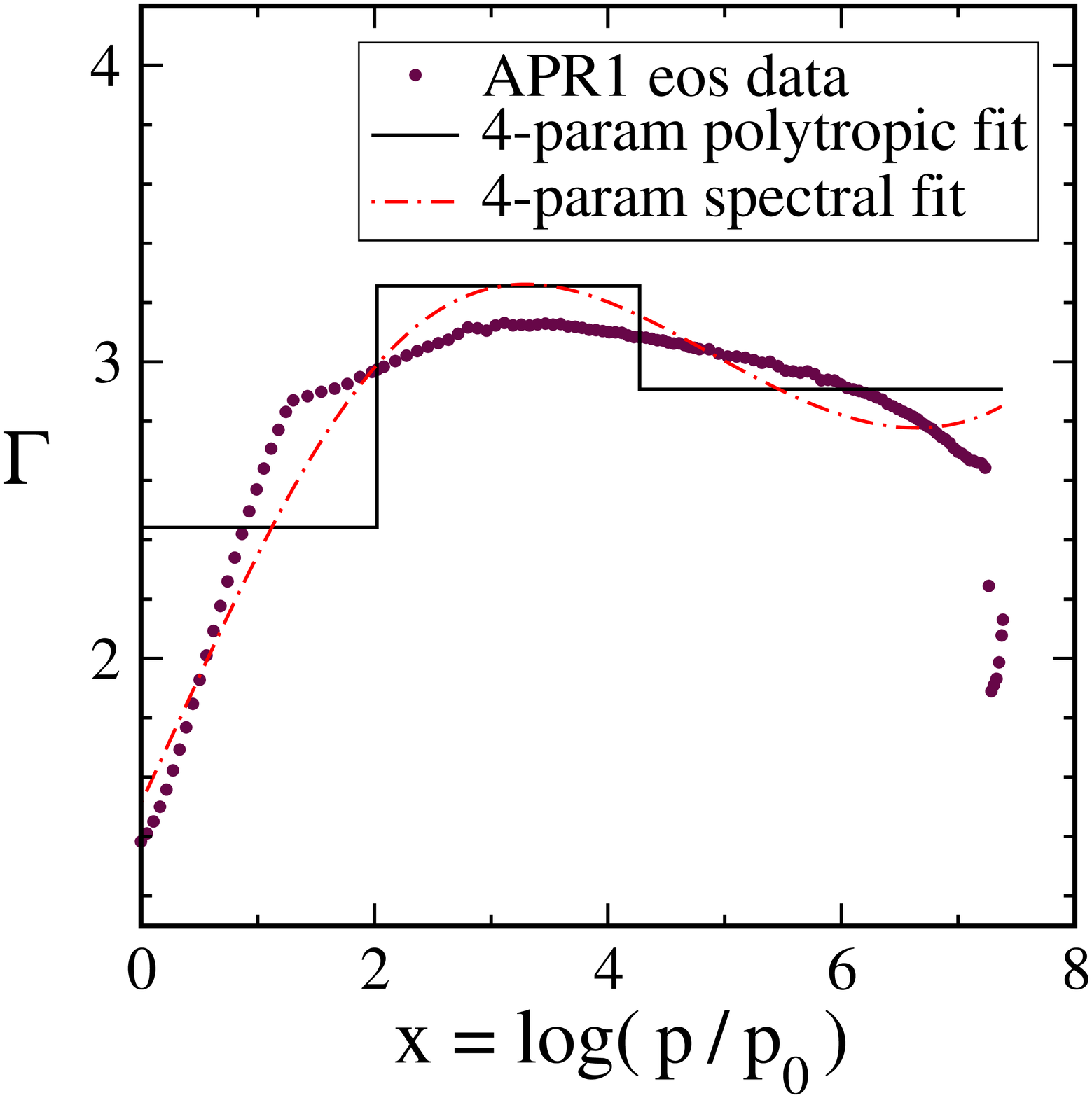}}
\caption{\label{f:apr1_gamma} Adiabatic index as a
function of pressure, $\Gamma(x)$, for various fits to the APR1
equation of state.}
\end{figure}
\begin{figure}[htbp!]
\centerline{\includegraphics[width=3in]{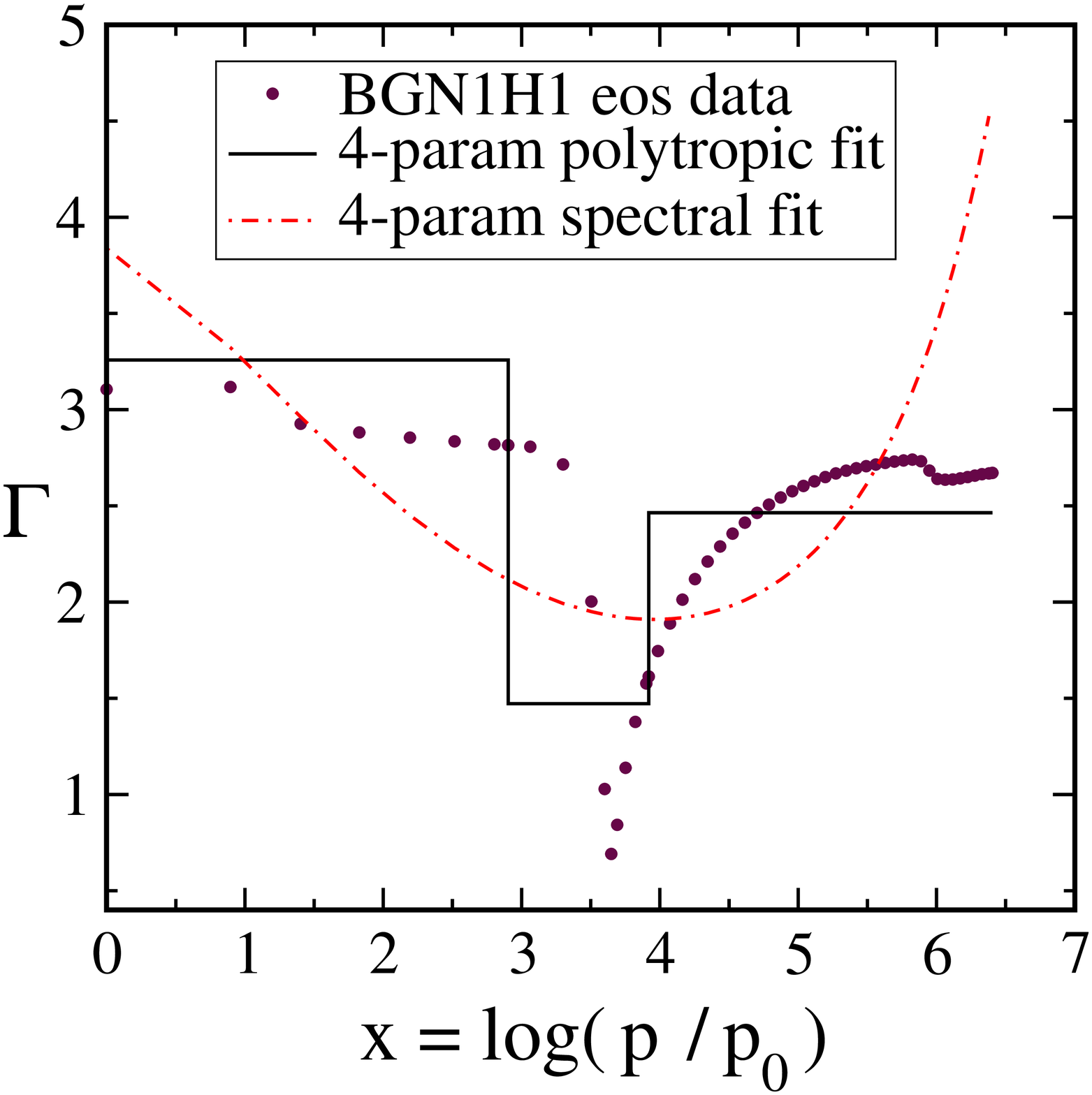}}
\caption{\label{f:bgn1h1_gamma} Adiabatic index as a function of
  pressure, $\Gamma(x)$, for various fits to the BGN1H1 equation of
  state.}
\end{figure}
\begin{figure}[htbp!]
\centerline{\includegraphics[width=3in]{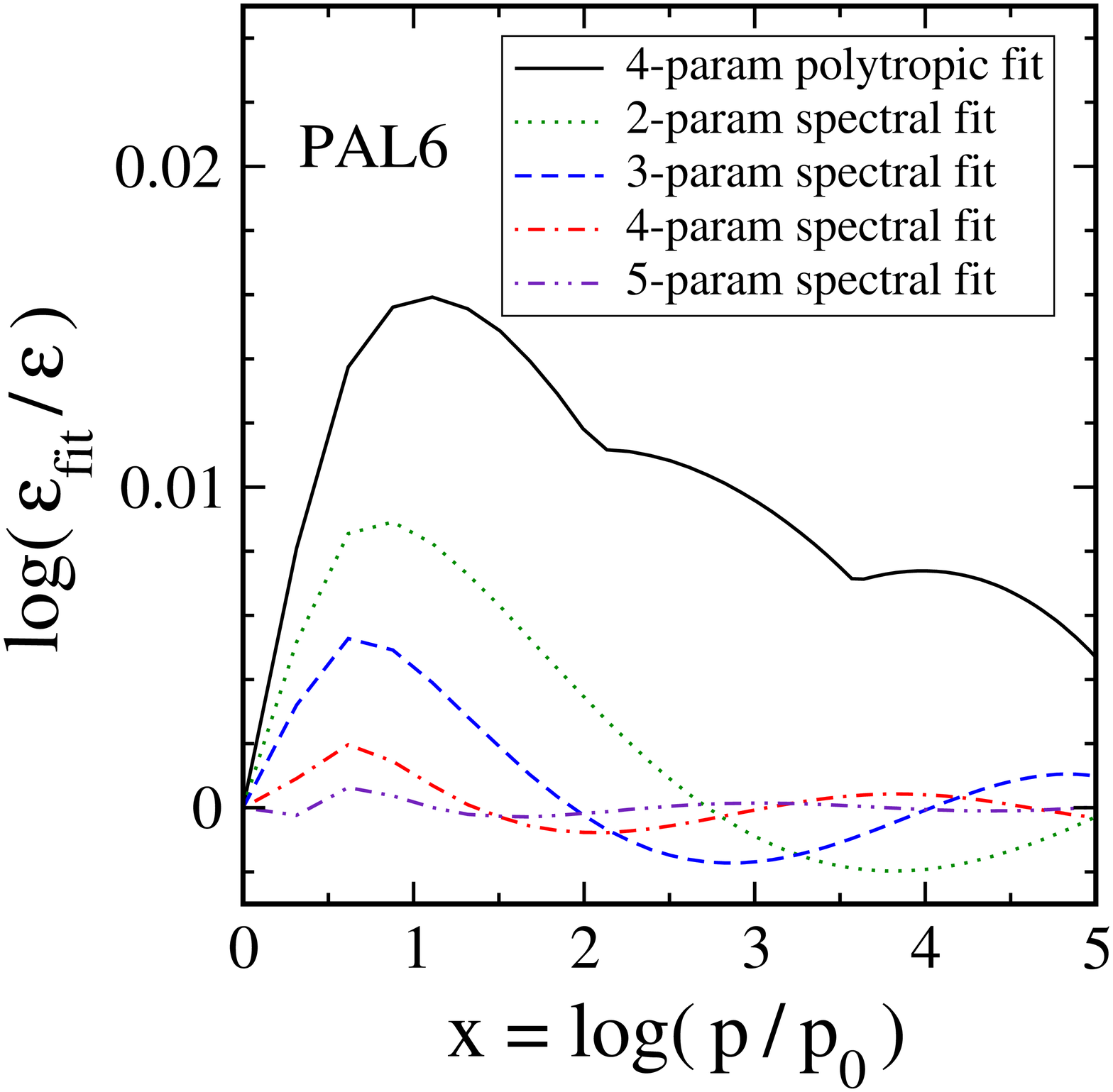}}
\caption{\label{f:pal6_comp_eos} Ratios between various fits,
  $\epsilon_{\mathrm{fit}}(x)$, and the exact PAL6 equation of state,
  $\epsilon(x)$.  Note that
  $\log(\epsilon_{\mathrm{fit}}/\epsilon)\approx
  (\epsilon_{\mathrm{fit}}-\epsilon)/\epsilon$ measures the fractional
  error of the fit.}
\end{figure}
\begin{figure}[t]
\centerline{\includegraphics[width=3in]{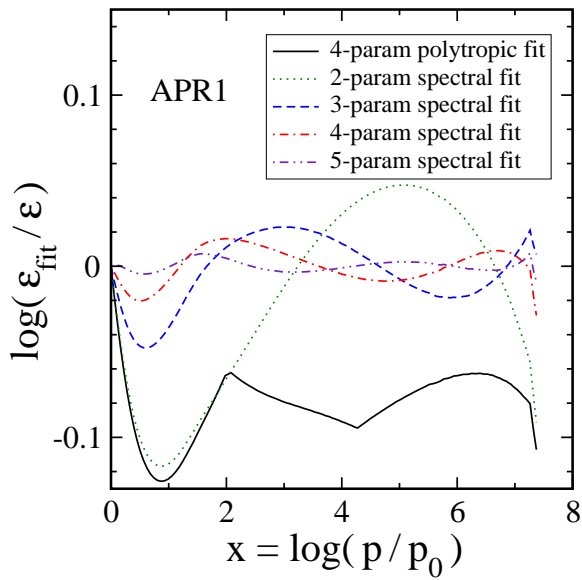}}
\caption{\label{f:apr1_comp_eos} Ratios between various fits,
  $\epsilon_{\mathrm{fit}}(x)$, and the exact APR1 equation of state,
  $\epsilon(x)$.  Note that
  $\log(\epsilon_{\mathrm{fit}}/\epsilon)\approx
  (\epsilon_{\mathrm{fit}}-\epsilon)/\epsilon$ measures the fractional
  error of the fit.}
\end{figure}
\begin{figure}[t]
\centerline{\includegraphics[width=3in]{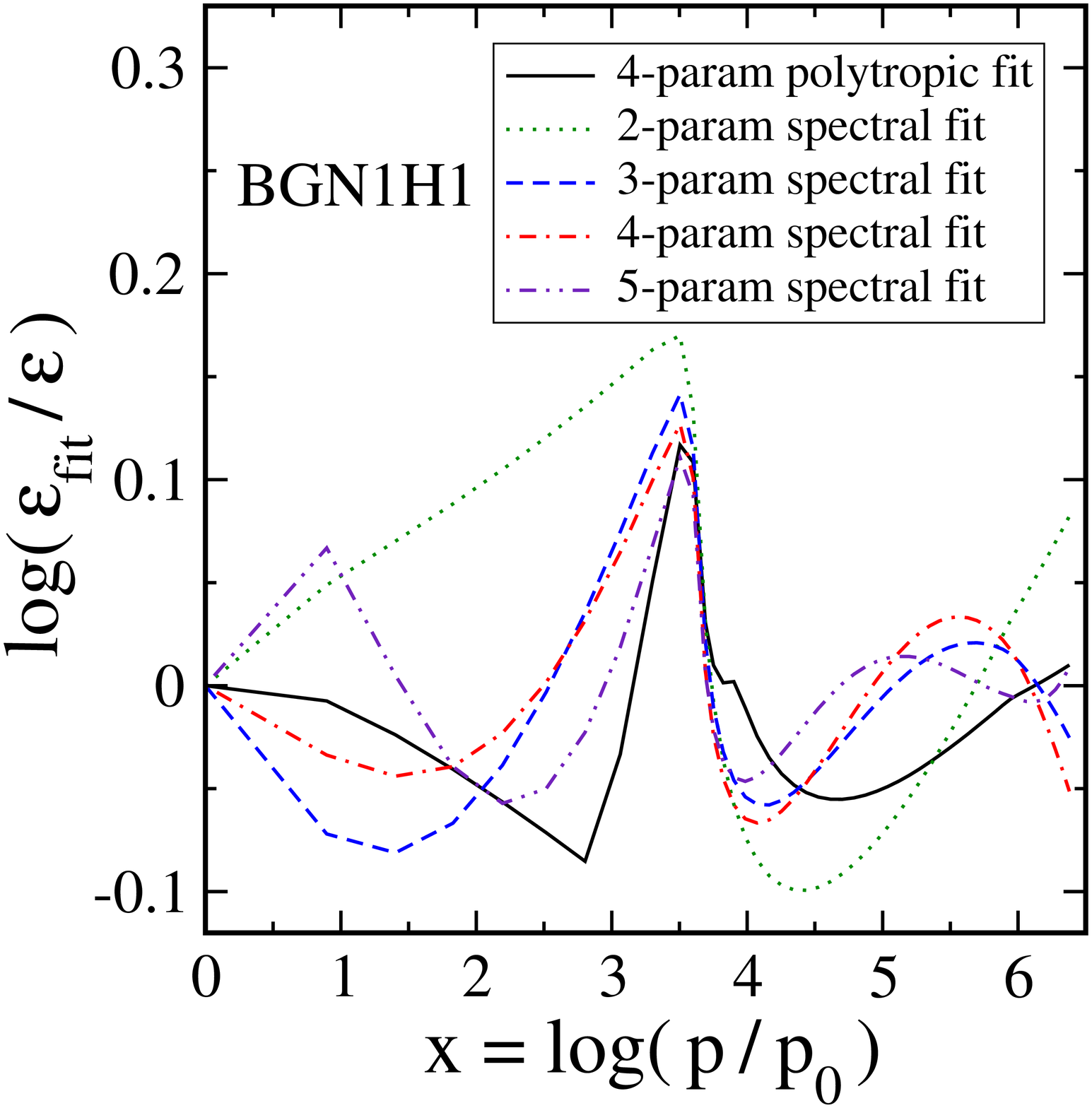}}
\caption{\label{f:bgn1h1_comp_eos} Ratios between various fits,
  $\epsilon_{\mathrm{fit}}(x)$, and the exact BGN1H1 equation of
  state, $\epsilon(x)$.  Note that
  $\log(\epsilon_{\mathrm{fit}}/\epsilon)\approx
  (\epsilon_{\mathrm{fit}}-\epsilon)/\epsilon$ measures the fractional
  error of the fit.}
\end{figure}

Three equations of state have been chosen from the complete set to
illustrate in more detail the accuracy of the fits.  These three cases
are equation of state PAL6 having the highest accuracy spectral fits,
APR1 having spectral fits with average accuracy, and BGN1H1 having the
lowest accuracy spectral fits.
Figures~\ref{f:pal6_gamma}-\ref{f:bgn1h1_gamma} show the adiabatic
index $\Gamma(x)$ computed directly from the tabulated equations of
state, the 4-parameter spectral fit to the adiabatic index, and the
4-parameter piecewise-polytrope fit for each of these equations of
state.  From these figures it is clear that the smoother equations of
state, like PAL6, have the highest accuracy fits, while the equations
of state with a sharp phase transition, like BGN1H1, have the lowest
accuracy fits.  Figures~\ref{f:pal6_comp_eos}-\ref{f:bgn1h1_comp_eos}
illustrate the errors in the various fits to the equation of state
$\epsilon(x)$ for these three example equations of state.  These
errors are illustrated as graphs of
$\log(\epsilon_{\mathrm{fit}}/\epsilon)\approx
(\epsilon_{\mathrm{fit}}-\epsilon)/\epsilon$, as functions of the
pressure variable $x=\log(p/p_0)$.  These figures show that the
spectral fits are much more accurate than the 4-parameter
piecewise-polytrope fit for the smooth and average equations of state,
PAL6 and APR1.  The spectral fits have about the same accuracy as the
piecewise-polytrope fit for equation of state BGN1H1 which has a
strong phase transition.
Figures~\ref{f:pal6_comp_eos}-\ref{f:bgn1h1_comp_eos} also illustrate
in a visual way the convergence of the spectral fits as the number of
basis functions is increased.
\begin{figure}[htb!]
\centerline{\includegraphics[width=3in]{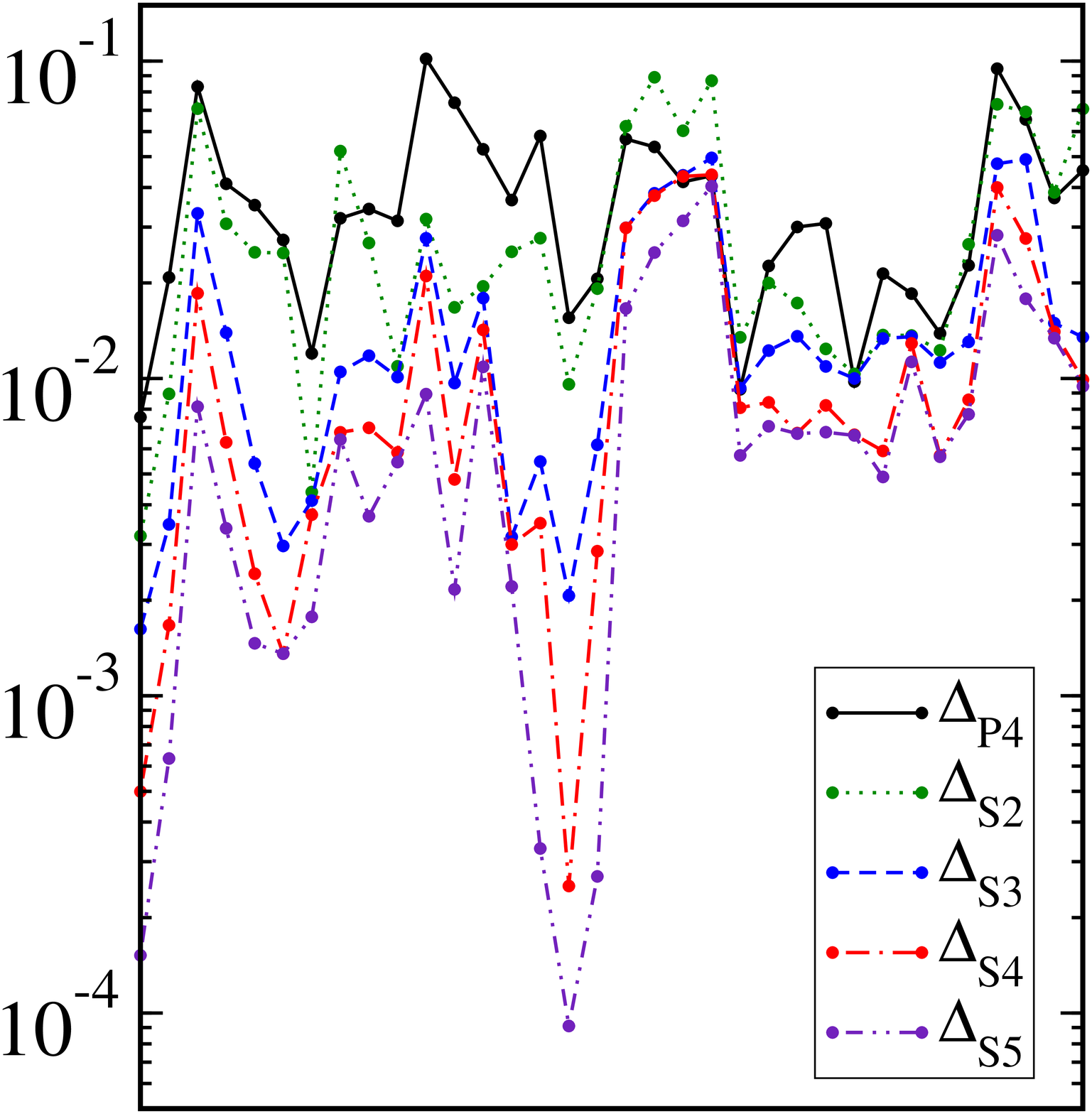}}
\caption{\label{f:chisqr_h} Residuals $\Delta_{Sk}$ are illustrated
  for several spectral fits and for the polynomial fit $\Delta_{P4}$
  for each of the realistic equation of state models, which are
  represented as points along the horizontal axis in this figure.
  These residuals are for enthalpy-based representations of the
  equation of state, $\epsilon=\epsilon(h)$, which are also given in
  Table~\ref{t:TableII}.}
\end{figure}
The enthalpy based representation of the equation of state,
$\epsilon=\epsilon(h)$ and $p=p(h)$, is more useful for certain
purposes.  It is helpful to understand, therefore, whether the
spectral fits for this representation,
Eqs.~(\ref{e:GammaH})--(\ref{e:TildeMuDef}), are as accurate
and effective as those for the standard $\epsilon=\epsilon(p)$
representation.  So approximate equations of state,
$\epsilon_{\mathrm{fit}}(h)$, have been constructed for the same set
of 34 realistic neutron star equations of state described in
Ref.~\cite{Read:2008iy}.  As before these fits were made by adjusting
the values of the spectral coefficients $\gamma_k$ defined in
Eq.~(\ref{e:GammaH}), to minimize the residual $\Delta(\gamma_k)$
defined in (\ref{e:ResidualDef}).  The only differences between this
and the standard case are: The variable $x=\log(h/h_0)$ is chosen here
to be an enthalpy variable, and the functions $\epsilon(h,\gamma_k)$
and $p(h,\gamma_k)$ are determined here with
Eqs.~(\ref{e:PressueH})--(\ref{e:TildeMuDef}).  Table~\ref{t:TableII}
lists each equation of state, along with the residuals, $\Delta_{S2}$,
$\Delta_{S3}$, $\Delta_{S4}$, and $\Delta_{S4}$, for the spectral fits
having 2, 3, 4 and 5 non-zero spectral coefficients.
The values of the residuals, $\Delta_{S2}$, $\Delta_{S3}$, 
$\Delta_{S4}$, $\Delta_{S5}$, and $\Delta_{P4}$, are also show
graphically in Fig.~\ref{f:chisqr_h} for these enthalpy based fits.
Table~\ref{t:TableII} also lists the coefficients, $\gamma_0$,
$\gamma_1$, $\gamma_2$, and $\gamma_3$ for the 4-parameter spectral fit.
Finally, Table~\ref{t:TableII} contains information about the enthalpy
variables, $h_0$ and $x_{\mathrm{max}}=\log(h_{\mathrm{max}}/h_0)$,
for each equation of state.  The quantity $h_0$ is the enthalpy for
which $\epsilon_0=\epsilon(h_0)$ and $p_0=p(h_0)$ whose values are
listed in Table~\ref{t:TableI}, and $h_{\mathrm{max}}$ is the value
for which $p_{\mathrm{max}}=p(h_{\mathrm{max}})$.  These results show
that the spectral expansions of the enthalpy based representation of
the equation of state are very comparable to the standard expansions.
\begin{table*}[!htb]
\begin{center}
\caption{Spectral Expansions of the $\epsilon=\epsilon(h)$ Form of
  Realistic Neutron-Star Equations of State
\label{t:TableII}}
\begin{tabular}{|l|cccc|cccc|cc|}
\hline\hline
    EOS   &$\Delta_{S2}$ &$\Delta_{S3}$ &$\Delta_{S4}$ 
     &$\Delta_{S5}$ &$\gamma_0$ &$\gamma_1$
    &$\gamma_2$ &$\gamma_3$ &$h_0$ &$x_{\mathrm{max}}$\\
\hline
    PAL6 &  0.0032 &  0.0016 &  0.0005 &  0.0002 &  0.8608 & -0.1509 &  0.0909 & -0.0192 &  0.0405 &    2.62\\
     SLy &  0.0089 &  0.0035 &  0.0017 &  0.0006 &  1.0077 &  0.2084 & -0.1266 &  0.0203 &  0.0314 &    3.22\\
     AP1 &  0.0708 &  0.0331 &  0.0186 &  0.0081 &  0.5205 &  1.3288 & -0.7804 &  0.1316 &  0.0276 &    3.31\\
     AP2 &  0.0307 &  0.0139 &  0.0063 &  0.0034 &  0.7557 &  0.6423 & -0.3586 &  0.0604 &  0.0287 &    3.34\\
     AP3 &  0.0250 &  0.0054 &  0.0024 &  0.0015 &  0.9520 &  0.4691 & -0.2109 &  0.0273 &  0.0298 &    3.40\\
     AP4 &  0.0249 &  0.0030 &  0.0014 &  0.0014 &  0.8824 &  0.4064 & -0.1405 &  0.0135 &  0.0297 &    3.41\\
     FPS &  0.0044 &  0.0041 &  0.0037 &  0.0018 &  1.1462 & -0.1088 &  0.0432 & -0.0106 &  0.0279 &    3.23\\
    WFF1 &  0.0521 &  0.0105 &  0.0068 &  0.0064 &  0.7115 &  0.8263 & -0.3258 &  0.0372 &  0.0273 &    3.53\\
    WFF2 &  0.0267 &  0.0118 &  0.0070 &  0.0037 &  0.8527 &  0.6162 & -0.3135 &  0.0512 &  0.0285 &    3.48\\
    WFF3 &  0.0109 &  0.0101 &  0.0058 &  0.0054 &  1.3660 & -0.4236 &  0.2293 & -0.0469 &  0.0253 &    3.30\\
    BBB2 &  0.0318 &  0.0277 &  0.0210 &  0.0089 &  0.8447 &  0.8101 & -0.5808 &  0.1120 &  0.0297 &    3.24\\
  BPAL12 &  0.0168 &  0.0097 &  0.0048 &  0.0022 &  1.0847 & -0.6537 &  0.4289 & -0.0874 &  0.0373 &    2.73\\
     ENG &  0.0195 &  0.0179 &  0.0142 &  0.0109 &  1.0426 &  0.4716 & -0.3353 &  0.0632 &  0.0286 &    3.39\\
    MPA1 &  0.0251 &  0.0032 &  0.0030 &  0.0022 &  1.0523 &  0.3546 & -0.1356 &  0.0074 &  0.0297 &    3.33\\
     MS1 &  0.0277 &  0.0055 &  0.0035 &  0.0003 &  0.9340 &  0.2231 &  0.0718 & -0.0642 &  0.0485 &    2.73\\
     MS2 &  0.0096 &  0.0021 &  0.0003 &  0.0001 &  0.9680 & -0.1326 &  0.0786 & -0.0360 &  0.0367 &    2.56\\
    MS1b &  0.0192 &  0.0062 &  0.0029 &  0.0003 &  1.2148 & -0.1255 &  0.1959 & -0.0700 &  0.0403 &    2.92\\
      PS &  0.0624 &  0.0298 &  0.0298 &  0.0166 &  1.3064 & -1.2479 &  0.5534 & -0.0186 &  0.0571 &    2.26\\
     GS1 &  0.0889 &  0.0383 &  0.0377 &  0.0249 &  1.7649 & -1.9609 &  0.8871 & -0.0942 &  0.0265 &    3.22\\
     GS2 &  0.0603 &  0.0437 &  0.0433 &  0.0314 &  1.2562 & -0.4936 & -0.1125 &  0.1051 &  0.0362 &    2.61\\
  BGN1H1 &  0.0868 &  0.0495 &  0.0439 &  0.0403 &  1.1915 &  0.2463 & -0.7222 &  0.2300 &  0.0356 &    2.97\\
    GNH3 &  0.0135 &  0.0093 &  0.0081 &  0.0057 &  1.0236 &  0.0297 & -0.2197 &  0.0742 &  0.0482 &    2.53\\
      H1 &  0.0200 &  0.0122 &  0.0084 &  0.0071 &  1.0365 &  0.1735 & -0.5519 &  0.1790 &  0.0406 &    2.39\\
      H2 &  0.0173 &  0.0136 &  0.0067 &  0.0067 &  1.0666 &  0.4639 & -0.8300 &  0.2393 &  0.0391 &    2.41\\
      H3 &  0.0124 &  0.0109 &  0.0082 &  0.0068 &  1.1204 &  0.2084 & -0.5285 &  0.1519 &  0.0402 &    2.42\\
      H4 &  0.0103 &  0.0100 &  0.0066 &  0.0066 &  1.0070 &  0.4040 & -0.3709 &  0.0696 &  0.0223 &    3.18\\
      H5 &  0.0137 &  0.0134 &  0.0059 &  0.0049 &  0.9614 &  0.6257 & -0.6459 &  0.1371 &  0.0233 &    2.96\\
      H6 &  0.0137 &  0.0135 &  0.0129 &  0.0113 &  1.0182 &  0.1483 & -0.2638 &  0.0628 &  0.0263 &    2.79\\
      H7 &  0.0123 &  0.0112 &  0.0057 &  0.0057 &  0.9147 &  0.4331 & -0.4779 &  0.1064 &  0.0254 &    2.94\\
    PCL2 &  0.0265 &  0.0130 &  0.0086 &  0.0077 &  1.0127 &  0.1289 & -0.4822 &  0.1616 &  0.0389 &    2.52\\
    ALF1 &  0.0731 &  0.0475 &  0.0400 &  0.0283 &  0.9349 & -0.3810 &  0.6697 & -0.2305 &  0.0305 &    2.75\\
    ALF2 &  0.0692 &  0.0490 &  0.0276 &  0.0178 &  0.7100 &  2.6577 & -2.2438 &  0.4712 &  0.0305 &    2.83\\
    ALF3 &  0.0386 &  0.0149 &  0.0140 &  0.0134 &  0.8987 &  0.4588 & -0.3533 &  0.0489 &  0.0305 &    2.73\\
    ALF4 &  0.0708 &  0.0135 &  0.0099 &  0.0094 &  0.8705 &  0.4469 &  0.0284 & -0.0914 &  0.0305 &    2.83\\
\hline
 Average  &0.0323  &0.0166  &0.0124  &0.0089&&&&&&\\
\hline\hline
 \end{tabular}
\end{center}
\end{table*}

\section{Discussion}

The spectral fits constructed here were designed to explore how
accurately the real neutron-star equation of state might be determined
once the first few accurate neutron-star mass-radius measurements
become available.  These fits could be improved for equations of state
with phase transitions by using separate spectral fits above and below
the phase-transition pressure.  These piecewise spectral fits could
eliminate Gibbs phenomena errors, but would require significantly more
(roughly double the number of) parameters.  Measuring this larger
number of parameters from neutron-star observations would therefore
require a much larger number of accurate mass-radius determinations.
A systematic study of the relative accuracy of single versus piecewise
spectral expansions (with the same total number of free parameters)
will therefore be delayed until more neutron-star measurements become
available.

The fits constructed here show that spectral representations of the
various realistic neutron-star equations of state are remarkably
accurate, even when the number of basis functions used in the spectral
expansion is rather small.  Such expansions provide an attractive
alternative to the piecewise-polytrope approximations for a variety of
current ``realistic'' models of the equation of state.  If the real
neutron-star equation of state is relatively smooth, then these
spectral expansions will provide an extremely efficient way to
represent it.  And remarkably, even if the neutron-star equation of
state has a phase transition, these spectral fits do about as well as
the piecewise polytrope fits with the same number of free parameters.


\acknowledgments I thank Benjamin Owen, Benjamin Lackey, John
Friedman, and Chris Vuille for helpful discussions about their work,
B.O., B.L. and C.V. for useful comments on an earlier draft of this paper,
and B.L and J.F. for providing tables of the various realistic
equations of state used here.  This research was supported in part by
a grant from the Sherman Fairchild Foundation, by NSF grant
PHY-0601459, and by NASA grant NNX09AF97G.


\vfill\eject
\bibstyle{prd} 
\bibliography{../References/References}
\end{document}